\documentclass[12pt]{article}

\usepackage{psfig}
\usepackage{amsfonts}

\newlength{\figwidth}
\newcommand{\eabe} {\begin{eqnarray}}
\newcommand{\eaen} {\end{eqnarray}}
\newcommand{\eqbe} {\begin{equation}}
\newcommand{\eqen} {\end{equation}}
\newcommand{\bibl}[5]
	{#1, {\it #2} {\bf #3} (#4) #5}

\newcommand{\anti}[1] {${ \ol \mrm #1 }$}
\newcommand{\pair}[1] {${\mrm {#1 \ol #1} }$}
\newcommand{\nmax} {n\srm{max}}
\newcommand{\Wmax} {W\srm{max}}
\newcommand{\shat} {s_1}
\newcommand{\mrm} {\mathrm}
\newcommand{\srm}[1] {_{\mathrm{#1}}}
\newcommand{\sfog} {s_2}

\newcommand{\ol} {\overline}
\newcommand{\aal} {{a_{\alpha}}}
\newcommand{\abe} {{a_{\beta}}}

\begin{document}

\begin{titlepage}
\begin{flushright}
 NORDITA 2000/36 HE\\
 April 2000
\end{flushright}
\vspace{25mm}
\begin{center}
  \Large
  {\bf On Energy Conservation in Lund String Fragmentation} \\
  \vspace{12mm}
  \normalsize
  Patrik Ed\'en\footnote{e-mail eden@nordita.dk}\\
  NORDITA\\
  Blegdamsvej 17, DK-2100 Copenhagen, Denmark\\
\end{center}
\vspace{5cm}
{\bf Abstract:} \\
I discuss a Monte Carlo algorithm for hadronization in the Lund string fragmentation picture, which conserves energy and momentum exactly as is proposed by the model. An embryo to this Monte Carlo is used to calculate the total decay density of a string. This is found to reach its predicted asymptotic behaviour already at a string energy of about six hadron masses.   
\end{titlepage}

\section{Introduction}
In the Lund string fragmentation model~\cite{string}, the colour field between a quark and an antiquark is assumed to be compressed into a 1+1 dimensional string, as the magnetic field in a super conductor. The density for a hadronic final state is a phase space factor times an exponential suppression of the space-time area swept by the fragmenting string. The model can be formulated in terms of an iterative fragmentation process, which makes it very suitable for MC implementation. A widely used MC program, \textsc{Jetset}\cite{jetset}, is based on the model.

As global energy--momentum conservation is ill suited for iterative algorithms, it is observed by additional approximations in the MC. In general, these approximations have negligible effects, but for subtle enough observables, and investigations at moderate energies, it may be worthwhile to point out that there is a difference between predictions of the Lund {\em model} and the Lund {\em Monte Carlo}.

Due to its success in describing data, the model is being used for more and more detailed investigations. 
One important example is Bose--Einstein correlation investigations at LEP, where one appealing model is based on the Lund string area law~\cite{BoBE}. Another possibility is to study of transverse momentum properties in pencil-like two-jet events, which due to the large LEP1 statistics can be selected without losing statistical significance. Different model predictions for these events are discussed in~\cite{twojet} and awaiting confrontation with data.

In experiments like the B-factories, JLAB and BEPC/BES in Beijing, it is possible to do detailed QCD studies at moderate energies. The more moderate is the energy, the more important is global momentum conservation treatment in models and Monte Carlo.

Examples of model vs.\ Monte Carlo differences, relevant at moderate energy few-body production, are given in~\cite{Hu}, which also presents a method to more precisely generate few-body states. A similar approach is used in~\cite{Maul}, to calculate (\pair q$\to \pi^+\pi^-)/($\pair q$\to$ hadrons), for moderate-energy \pair q systems.

In this paper I present a general algorithm, applicable to any multiplicity, which conserves energy and momentum exactly as proposed by the model. A simple embryo of this algorithm has been implemented~\cite{code}, and I use that to present numerical calculations of the total string decay density.

A set of extensions to the algorithm are needed, before a realistic generation of hadronic final states are possible. Most of these are postponed to future work, and only briefly discussed in the outlook. The results in this paper are obtained for the simple case of a 1+1 dimensional string decaying into hadrons of one flavour only. However, the algorithm is presented including two important extensions, namely several flavours and non-perturbative generation of transverse momenta.

The outline of this paper is as follows: In section~\ref{sec:model} the Lund string fragmentation model is reviewed. Section~\ref{sec:behov} deals with differences between Monte Carlo and model, and the new Monte Carlo algorithm is presented in section~\ref{sec:enmeson}. The calculation of the string decay density, based on this new algorithm, is presented in section~\ref{sec:g} and the paper ends with a summary and outlook in section~\ref{sec:summary}.

\section{The String Fragmentation Model}\label{sec:model}
\begin{figure}[tb]
\begin{center}
	\mbox{\psfig{figure=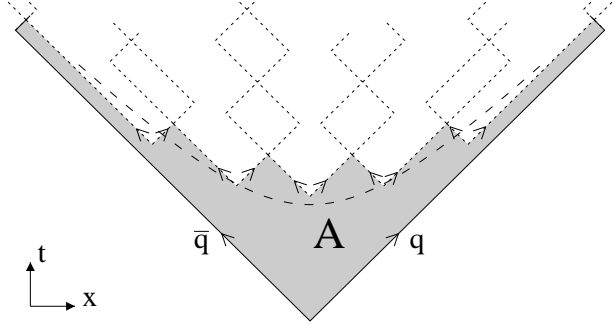,width=0.5\figwidth}}
\end{center}
\caption{\em $x-t$ picture of a fragmenting 1+1 dimensional string. A quark and an anti-quark from adjacent vertices form a meson (dotted lines). In the Lund model, the final state is exponentially suppressed in the colour coherence area $A$ (shaded region). The vertices tend to lie around a hyperbola of constant proper time (dashed line).}
\label{f:Adef}
\end{figure}
In this review, we start with the simple case of only one hadron flavour and no $p_\perp$.
The density for a string with mass $\sqrt{s}$ to produce $n$ hadrons is in the model given by
\eqbe g_n(s) \equiv \left[\prod_{r=1}^n N\int\mrm d^2p_r\delta(p_r^2 - m^2)\right]\delta^2(P-\sum_rp_r)\exp(-bA),~~~P^2=s. \label{e:gnsdef} \eqen
Here $N$ and $b$ are free parameters of the model, and $A$ is (the string tension squared times) the space--time area swept by the string, cf.\ Fig~\ref{f:Adef}. The density for a specific $n$-body state is given by the integrand of $g_n$. The index $r$ runs over the hadrons in {\em rank order}, specified so that the rank 1 hadron contains the original quark and an anti-quark \anti q$_1$, the rank 2 hadron contains the quark q$_1$ from the same break-up and an anti-quark \anti q$_2$ from the next vertex, etc.

The total decay density
\eqbe g(s) \equiv \sum_n g_n(s) \label{e:gdef} \eqen
can be written as an integral over a production factor for the rank 1 hadron times the decay density of the remaining string.
For $s \gg m^2$ the solution is of the form
\eqbe g(s) \propto s^a \label{e:gasympt}, \eqen
where $a$ is related to the parameters $N$ and $bm^2$ through
\eqbe 1 = N\int_0^1\frac{\mrm dz}{z}(1-z)^ae^{-bm^2/z}. \label{e:Nmdef} \eqen
$z$ represents the light-cone momentum fraction of the rank 1 hadron, which thus can be picked from a distribution
\eqbe f(z,m^2){\mrm d}z = N\frac{{\mrm d}z}z(1-z)^a\exp(-\frac{bm^2}{z}) \label{e:flund}. \eqen
This expression is independent of the string energy, which makes the formalism suitable for an iterative algorithm. Each produced hadron can be seen as the rank 1 hadron of a specific string remainder, and its momentum can be determined from Eq~(\ref{e:flund}). Eq~(\ref{e:Nmdef}) then acts as a unitarity constraint. This holds as long as the remaining energy is large enough for $g$ to have its asymptotic form in Eq~(\ref{e:gasympt}), when global energy--momentum conservation effects can be neglected.

The fragmentation function Eq~(\ref{e:flund}) gives left-right symmetry, i.e.\ we get the same final state distributions if we fragment iteratively from the quark end as from the anti-quark end of the string. Furthermore, the proper time distribution of a vertex is independent of rank, which implies that the vertices tend to be aligned along a hyperbola in $x-t$ space, cf.\ Fig~\ref{f:Adef}.

It is instructive (and historically more correct) to reverse the line of arguments. Searching for a left-right symmetric fragmentation function suitable for an iterative scheme, Eq~(\ref{e:flund}) can be shown to be a unique solution. The final state density, $G_n$, for an iteratively generated $n$-particle cluster
with momentum fraction $Z$ and mass $\sqrt{\shat}$ is then
\eqbe G_n(\shat,Z)\mrm d\shat\mrm dZ \equiv g_n(\shat) \mrm d\shat \frac{\mrm dZ}Z(1-Z)^a\exp(-b\shat \frac{1-Z}Z). \label{e:GnsZ} \eqen
This density  can be separated into one internal factor, independent of $Z$, and one external, depending only on $Z$ and ${\shat}$. The conventional choice is to let the internal part of the density be given by $g_n(\shat)$, which then is to be interpreted as the final state density of a fragmenting string with mass $\sqrt{\shat}$.

When we extend the model to several flavours, a set of new parameters enter. The inclusive probabilities $v_\alpha$ that a quark has flavour $\alpha$ and the probabilities $G^i_{\alpha\beta}$ that a quark $\alpha$ and antiquark $\beta$ form meson (or anti-meson) $i$ are free parameters which satisfy the constraints
\eqbe \sum_\alpha v_\alpha=1,~~~~\sum_i G^i_{\alpha\beta}=1,~~G^i_{\alpha\beta}=G^i_{\beta\alpha}. \label{e:smaknorm} \eqen
It is also possible to have flavour specific parameters $a_\alpha$ in the fragmentation function, which introduces a flavour dependence for the proper time distribution of a vertex.

Assuming no flavour correlations between vertices, the left-symmetric fragmentation function, Eq~(\ref{e:flund}), generalizes to
\eqbe  v_\beta G^i_{\alpha\beta}f_{\alpha\beta}(z,m_i^2){\mrm d}z = v_\beta G^i_{\alpha\beta}N_{\alpha\beta}(m_i^2)\frac{{\mrm d}z}zz^{a_\alpha}\left(\frac{1-z}z\right)^{a_\beta}\exp(-\frac{bm_i^2}{z}), \label{e:fflavlund} \eqen
which specifies the probability that flavour $\alpha$ is followed by flavour $\beta$, forming a meson $i$ which takes momentum fraction $z$. The $N$ and $a$ parameters are related by the unitarity constraint $\int{\mrm d}z f (z,m^2) = 1$.
With the definition
\eqbe C_{\alpha\beta}(m^2) \equiv N_{\alpha\beta}(m^2)e^{-bm^2}\left[\frac{a_\beta!(bm^2)^{a_\alpha}}{a_\alpha!(bm^2)^{a_\beta}}\right]^{\frac12}, \label{e:Cflavdef} \eqen
which in Eq~(\ref{e:Cintegrals}) is shown to satisfy $C_{\alpha\beta} = C_{\beta\alpha}$, the density for an iteratively produced $n$-particle cluster starting with flavour $\alpha_0$ and ending with flavour $\beta$ is
\eqbe G_n^{(0,\beta)}(\shat,Z){\mrm d}Z{\mrm d}\shat = \frac{v_\beta f_{0\beta}(Z,\shat){\mrm d}Z}{C_{0\beta}(\shat)}g_n^{(0,\beta)}(\shat){\mrm d}\shat. \label{e:GnsZflav} \eqen
The internal part of the density is given by
\eqbe g_n^{(0,\beta)}(s) = \!\left[\prod_{r=1}^{n}\!\sum_{\alpha_r,k_r}\! v_rG_{r\!-\!1,r}^{k_r}C_{r\!-\!1,r}e^{bm_{r}^2}\!\!\int\!\!\mrm d^2p_r\delta(p_r^2-m_{r}^2)\!\left(\!\frac{p_{+r}P_-}{P_+p_{-r}}\!\right)^{\!\!\frac{a_{r\!-\!1}-a_r}2}\right]\!\delta^2(P\!-\!{\scriptstyle \sum} p_r)\frac{\delta_{\alpha_n,\beta}}{v_\beta}e^{-bA}, \label{e:gnflav} \eqen
which specifies the decay density of a string.

The separation in internal and external parts is not unique, a factor depending only on $\shat$, $\alpha_0$ and $\beta$ can be associated with both parts.
This introduces no arbitrariness to observables, depending on relative densities for different final states, as long as we adopt a {\em factorization anzats}, saying that each string, generated by parton-level physics, decays into exactly one hadronic final state. The density $g_n^{(0,\beta)}$ is always left-right symmetric in ``truly'' internal properties, i.e.\ quark flavours $\alpha_1 ... \alpha_{n-1}$ and hadron flavours and momenta. In this paper the internal part is chosen to be symmetric also in its endpoint indices, $g_n^{(\alpha,\beta)}=g_n^{(\beta,\alpha)}$ and to approach the simpler one-flavour formulas in the limit of equal $a_\alpha$-values and equal masses $m_i$. 

We note that different phenomenological assumptions can be made to estimate the parameters $v_\alpha$ and $G^i_{\alpha\beta}$, and that, in general, several of them can be related to each other. 
In the standard Lund model, the string break-up is described by a tunneling process~\cite{tunnel}, which can be used to estimate suppression factors for strangeness and heavier spin states~\cite{string}.
In the UCLA approach~\cite{UCLA}, the product $v_\beta G_{\alpha\beta}N_{\alpha\beta}$ is assumed to be proportional to spin-- and isospin factors only, and the relative production of different mesons thus depend strongly on the factor $\exp(-bm^2/z)$ in the fragmentation function. (Strictly speaking, this implies flavour correlations between neighbouring vertices, i.e.\ the $v_\alpha$-parameters generalize to a set $v_{\alpha\beta}$.) In this paper, there is no need to specify the flavour selection mechanism, and I will continue to use the general, though perhaps unnecessarily numerous, parameters $v_\alpha$ and $G^i_{\alpha\beta}$.

\section{MC and Model Differences}\label{sec:behov}
Here I discuss implications of the differences in energy--momentum conservation between the \textsc{Jetset} MC~\cite{jetset} and the Lund model, represented by Eq~(\ref{e:gnflav}).

\begin{figure}[tb]
\begin{center}
	\mbox{\psfig{figure=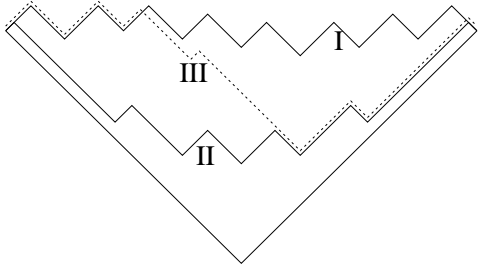,width=0.4\figwidth}}
\end{center}
\caption{\em Three final states. The densities for states $II$ and $III$ are determined by the same phase space factor, but state $III$ is further suppressed by a large string area. However, with the standard MC energy conservation recipe, merging of a type $I$ path from one string end and a type $II$ path from the other can occur without additional suppression, thus enhancing the occurrence of type $III$ states.}
\label{f:MCtrouble}
\end{figure}
In the iterative MC algorithm, hadrons are peeled off from the string ends until the remaining energy is getting small. Then two final hadrons are formed and their kinematics is essentially fixed by energy--momentum conservation. 
For 90 GeV strings, the approximate merging recipe affects two out of typically 10-15 primarily produced hadrons. For two-jet LEP events and experiments at more moderate energies, the final two particles constitutes a larger fraction of the primarily produced hadrons.

The degrees of freedom indicated by the vague expressions ``getting small'' and ``essentially fixed'' are carefully specified in the MC to give the two merging hadrons reasonable properties~\cite{jetset}. Nevertheless, there are some differences. Consider the three final states in Fig~\ref{f:MCtrouble}. The high-multiplicity state $I$ is enhanced by its phase-space factor but suppressed by a large area $A$. The low-multiplicity state $II$ has a small area, but is suppressed by a small phase space factor. The state $III$ is a permutation of $II$ and, therefore, has the same phase space factor. Its larger area makes it significantly more suppressed than state $II$.

Suppose that states $I$ and $II$ are equally probable. Letting both ends of the string fragment independently of each other implies that merging will occur between a $I$-type path from one side and a $II$-type path from the other just as often as between two paths of the same kind. This implies an enhanced occurrence of state $III$. Thus, the fluctuations in vertex proper times is slightly larger in the MC, and the correlation between rank and rapidity somewhat reduced.

In general, the MC and model differences are negligible. The bigger is the difference between state $II$ and $III$, the more suppressed are all the three discussed states. However, as mentioned in the introduction, this need not be the case at moderate energies well below the ${\mrm Z}^0$ peak~\cite{Hu}. At high energies (like 90GeV) the energy conservation procedure may be of relevance for subtle enough observables. Within the string fragmentation picture, it is possible to make different assumptions about e.g.\ non-perturbative $p_\perp$ generation~\cite{tunnel,jim,screw,twojet} and baryon production~\cite{popcorn,mops}, giving different predictions for correlations between flavours, $p_\perp$ and rank. As rank is not an observable, a confrontation with data must exploit the correlation between rank and rapidity. We note that the merging procedure in \textsc{Jetset}, where the correlation between rank and rapidity is reduced~\cite{Hu}, dilutes the observable differences between models for $p_\perp$-- and baryon generation.

\section{A new MC algorithm}\label{sec:enmeson}
In this section I derive the new algorithm, working in the case with several flavours. After presenting the algorithm, I spend one sub-section on some technical details related to the computer time consumption. The section ends with a sub-section describing extensions needed to include transverse momentum generation in string fragmentation.

We note that the decay density $g(s)$ is zero or a $\delta$-distribution for energies below the lowest two-particle threshold, cf.\ Eq~(\ref{e:gnsdef}). In the rest of this paper, I will constrain the discussion to energies above this threshold.

\subsection{1+1 dimensions}
\begin{figure}[tb]
\begin{center}
	\mbox{\psfig{figure=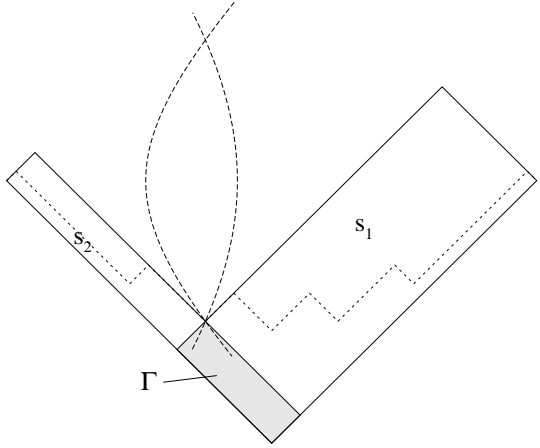,width=0.45\figwidth}}
\end{center}
\caption{\em Production of $n+m$ particles can be seen as production of two clusters. The hyperbolas in the figure show the possible end points for clusters with fixed mass squares $s_1$ and $s_2$. There are two solutions, where the hyperbolas meet. Each solution has its own $\Gamma$-value, corresponding to the proper time squared of the vertex common to the clusters.}
\label{f:Gdef}
\end{figure}
We rewrite $g_n$, Eq~(\ref{e:gnflav}), in the form
\eqbe \left\{\begin{array}{rcl} g^{(\alpha,\beta)}_1(s) & = & \displaystyle \sum_i G_{\alpha\beta}^i C_{\alpha\beta}(m_k^2) \delta(s-m_k^2) \\
 g^{(\alpha,\beta)}_{n+m}(s) & = & \displaystyle \sum_\gamma v_\gamma\int\mrm ds_1\mrm ds_2 \frac{g^{(\alpha,\gamma)}_n(s_1)g^{(\gamma,\beta)}_m(s_2)}{\sqrt{\lambda(s,s_1,s_2)}}\sum_{I=1}^2\Psi_{I}^{\alpha\beta,\gamma}(s,s_1,s_2) \end{array} \right.\eqen
where
\eqbe \lambda(a,b,c) \equiv a^2+b^2+c^2-2ab-2ac-2bc, \label{e:ldef} \eqen
\eqbe \Psi_{I}^{\alpha\beta,\gamma}(s,s_1,s_2) =  e^{-b\Gamma_I} \left(\frac{\sqrt{s_1s}}{s_1+\Gamma_I}\right)^{a_\alpha-a_\gamma}\left(\frac{\sqrt{s_2s}}{s_2+\Gamma_I}\right)^{a_\beta-a_\gamma}, \label{e:PsiIdef} \eqen
\eqbe \Gamma(a,b,c) \equiv \frac12\left(a-b-c\pm\sqrt{\lambda(a,b,c)}\right). \label{e:Gammadef} \eqen
$s_1$, $s_2$ and the two solutions to $\Gamma$ are illustrated in Fig~\ref{f:Gdef}.

The two-particle density $g_2$ is directly calculable, as is also all meson-specific two-particle densities  $g_{ij}$. For higher multiplicities we write
\eabe g^{(\alpha,\beta)}_{n+2}(s) & = & \sum_{\gamma,\delta, i,j} v_\gamma v_\delta G^i_{\gamma\delta} G^j_{\delta\beta} C_{\gamma\delta}(m_i^2) C_{\delta\beta}(m_j^2) \times \nonumber\\
&& \times
\int\frac{\mrm d\shat\mrm ds_2 g_n^{(\alpha,\gamma)}(\shat)}{\sqrt{\lambda(s,\shat,s_2)\lambda(s_2,m_i^2,m_j^2)}}\sum_{IJ}\Psi_I^{\alpha\beta,\gamma}(s,\shat,\sfog)\Psi_J^{\gamma\beta,\delta}(\sfog,m_j^2,m_i^2).\label{e:gn2flav} \eaen
Eq~(\ref{e:gn2flav}) is the basis of the algorithm. Before describing it in detail, I comment on a few things.

The cluster multiplicity is limited by $n+2 \le \sqrt s /m_{\mrm{min}}$. In general, the probability is negligible for much smaller $n$, since the typical multiplicity grows logarithmically with $s$. At this stage, it suffices to note that it {\em exists} an upper limit  $\nmax$ to $n$, which implies that we can select a preliminary $n$ from a flat distribution.

By ignoring the cluster momentum fraction $Z$, we can use iterative fragmentation to generate the mass and endpoint flavour of a $n$-particle cluster from the preliminary density (cf.\ Eq~(\ref{e:GnsZflav}))
\eqbe G_n^{(\alpha,\gamma)}(\shat){\mrm d}\shat \equiv \int_0^1{\mrm d}Z G_n^{(\alpha,\gamma)}(\shat,Z){\mrm d}\shat = \frac{v_\gamma}{C_{\alpha\gamma}(\shat)}g_n^{(\alpha,\gamma)}(\shat){\mrm d}\shat. \label{e:Gnsflav} \eqen

The denominator in Eq~(\ref{e:gn2flav}) goes like $[\sfog-(m_i+m_j)^2]^{\frac12}$ in the lower range of possible $\sfog$-values, and like $[(\sqrt s\! -\! \sqrt{\shat })^2-\sfog]^{\frac12}$ in the upper range. A preliminary distribution for $\sfog$, which takes these singularities into account and is suitable for MC generation, is
\eqbe \frac1{\pi}\mrm d\sfog\frac{\Theta(\sfog-(m_i+m_j)^2)\Theta((\sqrt s\! -\! \sqrt{\shat })^2-\sfog)}{\bigl[\sfog-(m_i+m_j)^2\bigr]^{\frac12}\bigl[(\sqrt s\! -\! \sqrt{\shat })^2-\sfog\bigr]^{\frac12}}. \label{e:s12prel}\eqen 
($\sfog$ is easily picked from this distribution by setting
	\eqbe \sfog = \frac12\left\{(\sqrt s\! -\! \sqrt{\shat })^2+(m_i+m_j)^2+\left[(\sqrt s\! -\! \sqrt{\shat})^2-(m_i+m_j)^2\right]\cos(R\pi)\right\}, \eqen
where $R$ is a random number from a uniform distribution between  0 and 1.)

After selecting  $n$, $\shat$ and $\sfog$ as mentioned, the set of four possible final states obtained should be accepted with a probability proportional to the weight
\eqbe W \equiv \pi\frac{C_{\alpha\gamma}(\shat)C_{\gamma\delta}(m_i^2)C_{\delta\beta}(m_j^2)}{\bigl[\sfog- (m_i-m_j)^2\bigr]^{\frac12}\bigl[(\sqrt s\! +\! \sqrt{\shat})^2-\sfog\bigr]^{\frac12}}\sum_{IJ}\Psi_I\Psi_J, \label{e:Wflavdef} \eqen 
This weight has a finite maximum $W\srm{max}$. The argument of $C$ is in the finite range $[m^2_{\mrm{min}},s]$, and there is thus and upper limit $C_{\mrm{max}}$. This implies that 
\eqbe W < \frac{\pi}{s^{\frac14}} \left[\frac{C_{\mrm {max}}}{\sqrt{m_{\mrm{min}}}}\left(\frac{\sqrt s}{m_{\mrm{min}}}\right)^{(a_{\mrm{max}}-a_{\mrm{min}})}\right]^3. \eqen
$\Wmax$ should be better estimated, to improve the acceptance of an event. That is discussed in the next subsection. At this stage, it suffices to note that it {\em exists} an upper limit  $\Wmax$, which implies that the acceptance probability is finite.

The algorithm, including two-particle production, is as follows:
\begin{itemize}
\item[1] Select $n$ with the probability 
\eqbe \left\{\begin{array}{ll}\frac{g_2}{(g_2+\Wmax\nmax)}, & n=0\\[2ex]\frac{\Wmax}{(g_2+\Wmax\nmax)}, & 1\le n\le \nmax\end{array}\right. . \eqen
\item[2a] If $n=0$ select hadrons $i$ and $j$ with probability $g^{(\alpha,\beta)}_{ij}(s)/g^{(\alpha,\beta)}_{2}(s)$ and then solution $I$. After that, the fragmentation is completed.
\item[2b] If $n>0$ iteratively generate $n$ particles with the symmetric Lund fragmentation function. Accept all momentum fractions $Z$ for the generated cluster. This creates a cluster with the mass squared $\shat$ and end quark $\gamma$ from the density in Eq~(\ref{e:Gnsflav}). 
\item[3] Select two final hadrons $i$ and $j$ with the probability $\sum_{\delta}v_\delta G_{\gamma\delta}^iG_{\delta\beta}^j$. If $\sqrt s <\sqrt{\shat}  + m_i + m_j$ return to point 1.
\item[4]  Select $\sfog$ from the preliminary distribution in Eq~(\ref{e:s12prel}).
\item[5] Accept the configuration with probability $W/W\srm{max}$. Else, return to point 1.
\item[6] Select solution $IJ$ with probability $\Psi_I\Psi_J/\sum_{IJ}\Psi_I\Psi_J$. Lorentz transform the cluster so that its last vertex is placed in the point specified by  $I$. Set the momenta of the final two hadrons according to  $J$.
\end{itemize}

\subsection{Computer Time Issues}\label{sec:time}
Above is presented an algorithm which, in finite time, produces a final state exactly according to the integrand in Eq~(\ref{e:gnflav}). Many extensions are needed before the algorithm can give a realistic description of hadronization. In particular, generalizing the algorithm to treat gluonic strings would require considerable effort. Before even considering such developments, it is relevant to discuss whether ``finite time'' also means ``reasonable time''. In this subsection I estimate the time consumption with the new algorithm, as compared to an algorithm where no normalization numbers $C$ need to be calculated, and where essentially each iteratively produced final state is accepted.

The normalization numbers $C$ must be numerically calculated, which in principle is a computer time problem. It is convenient to expand the integral representation of $C^{-1}$. The calculations are rather technical and not very illuminating, and here I only outline the procedure. For details, cf.~\cite{code}. 

The normalization number $C$ for a cluster with mass $\sqrt s$, defined in Eq~(\ref{e:Cflavdef}), can be written
\eabe C_{\alpha\beta}^{-1}(s)  & = & \sqrt{\frac{\aal!}{\abe!}}(bs)^{(\abe-\aal)/2}\int_0^1\frac{\mrm dz}zz^{\aal}(\frac{1-z}z)^\abe \exp(-bs\frac{1-z}z) = \nonumber \\ & = & \frac1{\sqrt{\aal!\abe!}(bs)^{1+\frac{\abe+\aal}2}}\int_0^\infty\mrm dx\mrm dy\, x^\aal y^\abe \exp(-x-y) \exp(-\frac{xy}{bs}). \label{e:Cintegrals} \eaen

Expanding $\exp(-\frac{xy}{bs})$ gives an asymptotic series which has to be truncated at some finite $N$,
\eabe C_{\alpha\beta}^{-1}(s) & \approx &  \frac1{\sqrt{\aal!\abe!}(bs)^{1+\frac{\abe+\aal}2}}\sum_{n=0}^N(-1)^n\frac{(n+\aal)!(n+\abe)!}{n!(bs)^n}. \label{e:highBsum} \eaen
As an example we get, for $\aal=\abe=0.5$ and $b= 0.75{\mrm{GeV}}^{-2}$, that this expansion can be used to obtain a precision $10^{-3}$ for all cluster masses above $\approx 4.3$GeV, and that it then suffices to include 13 terms in the expansion.

For small $bs$, it is possible to expand $(1-z)^\abe$ and do partial integrations to get an expansion in $bs$. In the example  $\aal=\abe=0.5$ and $b= 0.75{\mrm{GeV}}^{-2}$, it is enough to include approximately 70 terms in the $bs$ expansion, to obtain a precision  $10^{-3}$ in the whole range not covered by Eq~(\ref{e:highBsum}).

The sums in the $bs$ expansion contain large numbers of opposite signs which almost completely cancel each other. Thus, when very high precision is wanted, or when the $a$ values are very large (significantly larger than the reasonable upper limit $~$1), higher precision than in the computer floating point types is required on each term of the sum. In such a situation, a special floating point representation, where the precision can be set by the user, is required, which implies longer computation times. However, this is not a problem in the example with reasonable $a$-values and precision demands discussed here, and the numerical calculation of $C$ is not a big contributor to the computer time consumption.
 
The efficiency $E$, i.e.\ the average probability to accept an event, is in the new algorithm
\eqbe E\approx \frac{g(s)}{\nmax\Wmax}. \label{e:effdef} \eqen
This number has been investigated in the simple case of only one hadron. The result is $E\approx 1/30$ for 10 GeV strings. As discussed in section~\ref{sec:g}, there is little need to consider strings with higher energies.

This efficiency holds when numbers $\nmax$ and $\Wmax$, significantly better than the limits discussed after Eq~(\ref{e:gn2flav}), are found.  
At the start of a simulation, this is not the case, and computer time has to be spent on finding them. If identical strings are to be fragmented, the search for good $\nmax$ and $\Wmax$ need to be performed only once, but in general, when string configurations vary in different events, the calculation need to be done in every event. Results in the simple one-flavour case suggest that the final efficiency, including an empirical determination of $\nmax$ and $\Wmax$ in every event, is $E\approx 10^{-2}$.

The conclusion of this subsection is that the new fragmentation algorithm is expected to be roughly a factor 100 slower than the standard \textsc{Jetset} MC. We note that this time comparison applies only to the fragmentation phase. In a realistic event simulation, including parton cascades, resonance decays, detector simulation and event analysis routines, the relative time cost for the new algorithm will be substantially smaller.

\subsection{An algorithm with $p_\perp$}\label{sec:pt}
The model presented in section~\ref{sec:model} needs to be extended to provide a realistic description of hadron production. Most extensions are briefly discussed in the outlook, but here I present in detail an algorithm including non-perturbative generation of transverse momenta.

Adding transverse degrees of freedom implies that the two-particle density
$g_2$ no longer can be calculated exactly, but has to be generated by MC techniques. On the other hand, $g_2$ is no longer singular, which means that two-particle states can be treated on more equal footing with the higher multiplicities. Thus, this model extension significantly changes the algorithm, while many of the extra features discussed in the outlook should be more straightforward to implement.

The standard way to describe $k_\perp$ generation in the string model is via a tunneling mechanism~\cite{tunnel}, which gives rise to an azimuthally symmetric Gaussian distribution. Here I present an algorithm applicable when a quark and antiquark in a vertex gets balancing transverse momenta $\ol k_\perp$ from a distribution $T(\ol k_\perp){\mrm d}^2k_\perp$ without singularities, and where there is no correlation between different vertices.

Suppressing the $\ol k_\perp$ dependence in the notation, the equation relevant for the algorithm reads
\eqbe \frac{\sqrt{a_\alpha!}g_{n+2}^{(\alpha,\beta)}(s)}{(bs)^{\frac{a_\alpha-a_\beta}2}} \! = \! \sum_\gamma \!\int\!\mrm d^2\ol k_{\perp \gamma}\mrm dZ\mrm d\shat \frac{G_n^{(\alpha,\gamma)}(\shat,Z)}{(1-Z)^{a_\beta}}\frac{\sqrt{a_\gamma!}g_2^{(\gamma,\beta)}(\sfog)}{(b\sfog)^{\frac{a_\gamma-a_\beta}2}},\label{e:gn2spt} \eqen
where the density  $G_n^{(\alpha,\gamma)}(\shat,Z)\mrm d^2\ol k_{\perp \gamma}\mrm dZ\mrm d\shat$ is obtained by iterative fragmentation. This will be used to determine the cluster variables $\shat$ and $Z$, which implies that squared transverse mass of the 2-particle state is 
\eqbe \sfog = (1-Z)(s-\shat/Z). \label{e:s2def} \eqen

Guided by Eq~(\ref{e:gn2spt}), we can formulate an algorithm as follows:
\begin{itemize}
\item[1] Select $n$ between 0 and $\nmax$ from a uniform distribution.
\item[2] Generate iteratively $n$ particles with the symmetric Lund fragmentation function. Return to 1 if the remaining string energy is less than $2m\srm{min}$. (if $n=0$, no particles are generated. Instead set $Z=0$, $\sfog = s$ and $\gamma=\alpha$.) It now remains to form two hadrons from the string remainder build up by a quark $\gamma$ with transverse momentum $\ol k_{\perp\gamma}$ and an anti-quark $\beta$ with transverse momentum $-\ol k_{\perp\beta}$, and having a squared mass $\sfog$ given by Eq~(\ref{e:s2def}).
\item[3] Select flavour $\delta$ for the final vertex and flavours $i,j$ for the two final hadrons with probability $v_\delta G^i_{\gamma\delta}G^j_{\delta\beta}$. 
\item[4] Define perpendicular $x$ and $y$ axes so that $k_{y\gamma} - k_{y\beta} =0$. 
Select $k_x$ in the last vertex from the distribution $\left[\int{\mrm d}k_yT(\ol k_\perp)\right]{\mrm d}k_x$.
\item[5] Calculate $p_{xi}=k_{x\gamma}-k_x$ and $p_{xj}=k_x-k_{x\beta}$. If $\sqrt{\sfog}<\sqrt{m_i^2+p_{xi}^2}+\sqrt{m_j^2+p_{xj}^2}$ return to 1.
\item[6] Select a random number $R$ from a flat distribution between 0 and 1. Set
\eqbe k_y = k_{y\gamma} + \frac 12\sqrt{\frac{\lambda(\sfog,m_i^2+p_{xi}^2,m_j^2+p_{xj}^2)}{\sfog}}\cos(R\pi) \label{e:kyR} \eqen
\item[7] Accept the event with a weight
\eqbe W \propto \frac{T(\ol k_\perp)}{\int{\mrm d}k_yT(\ol k_\perp)}\frac{C_{\gamma\delta}(m_{\perp i}^2)  C_{\delta\beta}(m_{\perp j}^2)}{(1-Z)^{a_\beta}\sqrt{\sfog}}
\frac{\sqrt{a_\gamma!}}{(b\sfog)^{\frac{a_\gamma-a_\beta}2}}\sum_{I=1}^2\Psi_I^{\gamma\beta,\delta}(\sfog,m_{\perp i}^2,m_{\perp j}^2). \label{e:Wpt} \eqen
If not accepted, return to 1.
\item[8] Select solution $I$ and place the final two particles accordingly.
\end{itemize}

\section{The Total String Decay Density}\label{sec:g}
As discussed in section~\ref{sec:model}, observables depend on relative densities for different hadronic final states possible from a string, and not on its total decay density $g(s)$.
Nevertheless, a numerical calculation of $g(s)$ is of interest. It is instructive to see at what energies $g(s)$ --- in whatever form it has been specified --- approaches its asymptotic behaviour.
When the asymptotic limit is valid, the iterative algorithm, postponing the energy momentum conservation problem to a later stage, is a good approximation.

It is much simpler to calculate the densities $g_n(s)$, and thus their sum $g(s)$, than to generate final states. In the $1+1$ dimensional model without $p_\perp$ generation it can be done as follows: The two-particle density, $g_2(s)$, is directly calculable. For all higher multiplicities $n+2$, generate an $n$-particle cluster and calculate its mass $\sqrt{\shat}$. Set the ``merging weight'' $W=0$ if it is kinematically impossible to produce  a final state with a cluster of mass $\sqrt{\shat}$ and two more hadrons. Else, select a mass $\sqrt{\sfog}$ for the final two particles from the distribution in Eq~(\ref{e:s12prel}) and calculate the weight $W$ in Eq~(\ref{e:Wflavdef}). Repeating the procedure gives $g_n(s)$ as the average of $W$.

\begin{figure}[tb]
\begin{center}
	\mbox{\psfig{figure=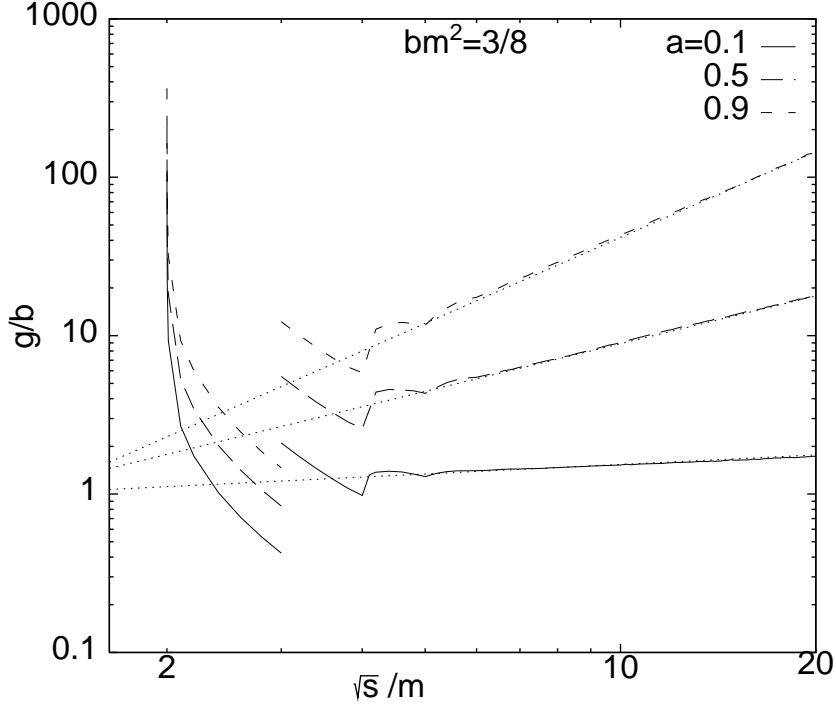,width=0.7\figwidth}}
\end{center}
\caption{\em The decay density $g(s)/b$ as a function of $\sqrt s/m$, for $bm^2=3/8$ and three different $a$ values. At $\sqrt s/m = 2,3,4$ clear threshold effects are seen, but already at $\sqrt s \sim 6 m$ the density has well approached its asymptotic form, indicated by dotted straight lines. Results for other values of $bm^2$, corresponding to meson masses in the range 0.5 to 1 GeV, are similar.}
\label{f:gtot}
\end{figure}
Fig~\ref{f:gtot} shows results in the simple model without $p_\perp$ and with only one meson flavour of mass $m$. The dimensionless quantity $g(s)/b$ is plotted as a function of $\sqrt s/m$, for different values of $a$. The quantity $bm^2$ is set to $3/8$, corresponding to a typical value $0.75$GeV$^{-2}$ on $b$ and typical mass $m=0.7$GeV. One can clearly see the threshold effects for $\sqrt s/m = 2, 3, 4$. After that, the contribution of $g_n(s=n^2m^2)$ is getting less important relative to lower multiplicity contributions, and the sum $g(s)$ is well approximated by its asymptotic form at $\sqrt s\sim 6m$. 

This implies that high energy strings may be iteratively fragmented using the symmetric fragmentation function, as long as there is a negligible risk to get a string remainder with an energy below some six hadron masses. In doing so, the algorithm efficiency, discussed in section~\ref{sec:time} for a string energy of about $15m$, does not deteriorate with increasing energy.

\section{Summary and Outlook}\label{sec:summary}
The Lund string fragmentation picture~\cite{string} is successful in comparison with data, and the computer implementation \textsc{Jetset}~\cite{jetset} is a widely used event generator for high energy physics experiments. However, there are differences between Lund String fragmentation model and Monte Carlo. Global energy momentum conservation does not fit into an iterative fragmentation scheme and is observed in a slightly different way in the MC, as compared to the model mathematical expressions. The differences may be non-negligible in certain cases, like few-body production~\cite{Hu} and subtle enough observables, as those discussed in~\cite{twojet}.

I present an algorithm which observes energy--momentum conservation exactly as proposed by the model expressions. An embryo to a computer program based on this algorithm~\cite{code} is used to calculate the total decay density $g(s)$ of a string. It is shown to be well approximated by its asymptotic behaviour for string energies above some six hadron masses.

A complete fragmentation simulator has not yet been developed, but only the parts needed to calculate $g(s)$. With this program embryo, it is possible to estimate the time consumption of a future event simulation program, which is found to be roughly 100 times larger than \textsc{Jetset}. It should be noted that this is only in the fragmentation phase. In a realistic calculation, including parton cascades, resonance decays, detector simulation and event analysis routines, the relative time cost for the new algorithm will be substantially smaller.

A set of extensions to the algorithm are needed, before a realistic generation of hadronic final states are possible.
Inclusion of flavour-- and $p_\perp$ correlations as in~\cite{jim}, baryon production as in~\cite{popcorn}, modifications to the area-definition in the case of heavy endpoint quarks~\cite{Bowler} and Breit-Wigner mass distributions for unstable resonances should all be straightforward developments, while more advanced baryon models~\cite{mops} and strings with gluons may need more effort. The latter is a particularly complex problem, as apparent from the thorough description in~\cite{gluons}.

\subsubsection*{Acknowledgments}
I thank prof.\ G\"osta Gustafson for stimulating discussions.


\begin{thebibliography}{99}
\bibitem{string}
  \bibl{B. Andersson, G. Gustafson, G. Ingelman, T. Sj\"ostrand}
	{Phys.\ Rep.} {97} {1983} {31}
\bibitem{jetset}
  \bibl{M. Bengtsson, T. Sj\"ostrand}{Comp.\ Phys.\ Comm.}{39}{1986}{347};\\
  \bibl{T. Sj\"ostrand}{Comp.\ Phys.\ Comm.}{82}{1994}{74}
\bibitem{BoBE}  
  \bibl{B. Andersson, W. Hofmann}{Phys.\ Lett.}{B169}{1986}{364};\\
  \bibl{B. Andersson, M. Ringn\'er}{Nucl.\ Phys.}{B513}{1998}{627}
\bibitem{twojet} 
  \bibl{P. Ed\'en, G. Gustafson}{Eur.\ Phys.\ J.}{C8}{1999}{435}
\bibitem{Hu}  B. Andersson, H. Hu, hep-ph/9910415 (1999)
\bibitem{Maul} M. Maul, hep-ph/0003254 (2000)
\bibitem{code} C$^{++}$ source code, available at http://nordita.dk/\~{}eden/gns, or on request from the author.
\bibitem{tunnel}
  \bibl{H. Bohr, H.B. Nielsen}{NBI-HE-78-3} {} {1978} {}
\bibitem{UCLA}
  \bibl{C.D. Buchanan, S.B. Chun}{Phys.\ Rev.\ Lett.}{59}{1987}{1997};\\
  \bibl{C.D. Buchanan, S.B. Chun}{Phys.\ Rep.}{292}{1998}{239}
\bibitem{jim}
  \bibl{B. Andersson, G. Gustafson, J. Samuelsson}{Z. Phys.}{C64}{1994}{653}
\bibitem{screw}
  \bibl{B. Andersson, G. Gustafson, J. H\"akkinen, M. Ringn\'er, P. Sutton}{JHEP}{98-09}{1998}{014} 
\bibitem{popcorn}
  \bibl{B. Andersson, G. Gustafson, T. Sj\"ostrand}{Phys. Scripta}{32}{1985}{574}
\bibitem{mops}
  \bibl{P. Ed\'en, G. Gustafson}{Z. Phys.}{C75}{1997}{41}
\bibitem{Bowler}
  \bibl{M.G.Bowler}{Z. Phys.}{C11}{1981}{169};\\
  \bibl{D.A.Morris}{Nucl. Phys.}{B313}{1989}{634}
\bibitem{gluons}
  \bibl{T. Sj\"ostrand}{Nucl.\ Phys.}{B248}{1984}{469}
\end{thebibliography}
\end{document}